\documentclass[aps,pre,twocolumn,superscriptaddress,showpacs,draft]{revtex4}
\input{epsf}
\usepackage{amsmath}
\usepackage{amsfonts}
\usepackage{dcolumn}                             
\usepackage{amssymb}
\usepackage{graphicx,graphics,wrapfig,rotating}         
\usepackage[german,english]{babel}
\usepackage{bm,fancybox}                         

\newcommand{\rem}[1]{}

\newcommand{\ket}[1]{|\,#1\,\rangle}                %
\newcommand{\bra}[1]{\langle\,#1\,}                 %
\newfont{\Bb}{msbm10}                   %
\newcommand{\tr}{\mbox{tr}}             %
\begin{document}

\title{Thermal effects on chaotic directed transport}

\author{Gabriel G. Carlo}
\affiliation{Departamento de F\'\i sica, CNEA, Libertador 8250, (C1429BNP) Buenos Aires, Argentina}
\author{Mar\'\i a E. Spina}
\affiliation{Departamento de F\'\i sica, CNEA, Libertador 8250,
(C1429BNP) Buenos Aires, Argentina}

\date{\today}


\begin{abstract}

We study a chaotic ratchet system under the influence of a thermal environment. By direct integration of the
Lindblad equation we are able to analyze its behavior for a wide range of couplings with the environment,
and for different finite temperatures. We observe that
the enhancement of the classical and quantum currents due to temperature depend strongly on the specific 
properties of the system. This makes difficult to extract universal behaviors. We have also found that 
there is an analogy between the effects of the classical thermal noise and those
of the finite $\hbar$ size. These results open many possibilities for their testing and implementation 
in kicked BECs and cold atoms experiments.

\end{abstract}

\maketitle

\section{Introduction}

The first theoretical constructions related with directed transport have been formulated in an early
work by Feynman \cite{Feynman}. This opened a field whose relevance and activity has
been increasing since then. The motivation is twofold,
in the first place several fundamental questions have originated from these ideas and have been
only partially answered \cite{Reimann}. As a consequence, the subject grew into a major
new field of statistical physics. On the other hand, the wealth of possible applications has determined
the emergence of a very broad area of research in physics. Ratchets, generically defined as periodic
systems having a dissipative dynamics associated with thermal noise and unbiased perturbations
(driving them out of equilibrium), can be used to model a wide range of different phenomena.
In order to give just a few examples we can
mention molecular motors in biology \cite{biology}, nanodevices (rectifiers, pumps, particle
separators, molecular switches and transistors) \cite{nanodevices}, and coupled
Josephson junctions \cite{Zapata}. On the other hand, there is a 
great interest in the theoretical description and
experimental implementations of cold atoms subjected to
time-dependent standing waves of light. They play a central role in many studies
of the quantum dynamics of nonlinear systems (dynamical localization, decoherence, quantum 
resonance, etc.) \cite{KickedCA}. In particular, the so-called optical ratchets, i.e. 
directed transport of laser cooled atoms, have been successfully implemented in this 
sort of experiments \cite{CAexp}.

The appearance of a net current can be classically explained by the necessary condition
of breaking all spatiotemporal symmetries leading to momentum inversion \cite{origin}.
This, along with Curie's principle indicates that transport should be present, given that it is
not forbidden. As an example of this situation we can mention Hamiltonian systems (with necessarily
mixed phase spaces) where a chaotic layer should be asymmetric \cite{Ham}. In the more
general dissipative case, chaotic attractors having this property are necessary \cite{nonHam}.
It is usual that the same principle translates almost directly into the quantum domain \cite{asymFloquet},
and very similar behaviors arise. But sometimes quantum mechanics introduces new effects
\cite{Qeffects} and the results depart from the classical ones. For example, the efficiency
of a forced thermal quantum ratchet has been calculated in \cite{Ghosh}. In that work the authors find
that the quantum current is higher in comparison with the classical one at
the lowest values of the temperature. As this parameter increases the discrepancies diminish and finally
they become negligibly small.

Recently, a quantum chaotic dissipative ratchet has been introduced in \cite{Carlo}.
In this example directed transport appears for
particles under the influence of a pulsed asymmetric potential in the presence of
a dissipative environment at zero temperature. The asymmetry of the quantum strange
attractor is at the origin of the quantum current, in close analogy with what happens
at the classical level. Indeed, this work provides with the parameters needed for a
possible implementation using cold atoms in an optical lattice. For a somewhat similar
dynamics, the case of weak coupling and low temperature has been studied
in \cite{DenisovR}. In the present work we extend the study of \cite{Carlo}
for a wide range of couplings with the environment and different temperature values.
We have verified that there is a strong dependence of the current behavior
on the coupling strength. If we compare this with the results found in \cite{Ghosh},
for instance, we could not find a generic enhancement of the quantum current for finite temperatures.
Instead of that, we could identify a close quantum-to-classical correspondence when
considering thermal effects only at the classical level. In fact, we have found that
the finite $\hbar$ effects on the quantum current are analogous
to the influence of the classical thermal fluctuations on the classical transport.

In the following we describe the organization of this paper. In Section II we present our
model for the system and for the environment, explaining the methods we have used to investigate
the current behavior. In Section III we show the results where the roles of $\hbar$, the coupling
strength and the temperature are analyzed in detail. Finally, in Section IV we
summarize and point out our conclusions.

\section{The system and the environment}

In this Section we describe the approach used to model the system plus the environment.
We study the motion of a particle in a periodic kicked asymmetric potential given by
\begin{equation}
V(x,t)=k[\cos{x} + a/2 \cos{(2x + \Phi)}] \sum_{m=-\infty}^{+ \infty}
\delta (t-m \tau)
\end{equation}
where $\tau$ is the kicking period, $k$ is the strength of the
kick, and $a$ and $\Phi$ are parameters that allow to introduce a
spatial asymmetry \cite{Carlo}. The effects of the environment are taken into
account by means of a velocity dependent damping and thermal
fluctuations. At the classical level, these ingredients are
incorporated in the following map:

\begin{equation}
\left\{
\begin{array}{l}
\overline{n}=\Gamma n + k[\sin{x} +a \sin{(2 x+ \Phi)}]+ \xi \\
\overline{x}= x + \tau \overline{n}
\end{array}
\right.
\end{equation}
In these expressions, $n$ is the momentum variable conjugated to
$x$ and $\Gamma$ is the dissipation parameter, with $ 0 \leq
\Gamma \leq 1$ . The thermal noise $\xi$ is related to $\Gamma$,
according to $ <\xi^2> =2 (1-\Gamma) k_B T$, where $k_B$ is the
Boltzmann constant and $T$ is the temperature, making the
formulation consistent with the fluctuation-dissipation
relationship. By performing the change of variables $\tau n
\rightarrow p$, $\tau k \rightarrow K$, and $\tau \xi \rightarrow
\tilde \xi$ (where $ <{\tilde \xi}^2> =2 (1-\Gamma) k_B \tilde{T}$
and $\tilde T = \tau^{2} T$), we can eliminate the period from the
classical expressions, and define the new map
\begin{equation}
\left\{
\begin{array}{l}
\overline{p}=\Gamma p + K[\sin{x} +a \sin{(2 x+ \Phi)}]+ \tilde \xi \\
\overline{x}= x + \overline{p}
\end{array}
\right.
\label{cl_map}
\end{equation}

In the quantum version of the model the system Hamiltonian is given by
\begin{equation}
\hat{H}_S = \hat{n}^2/2 + V(\hat{x},t)
\end{equation}
where the quantization has been performed in such a way \cite{Izrailev} that
$x \rightarrow \hat{x}$, $n \rightarrow \hat{n}=-i(d/dx)$ and
$\hbar=1$. This amounts to saying that, being $[\hat{x},\hat{p}]=i\tau$,
there is an effective Planck constant given by $\hbar_{\rm eff}=\tau$. Then,
the classical limit corresponds to $\hbar_{\rm eff} \rightarrow 0$, while
keeping $K=\hbar_{\rm eff} k$ constant.

In order to incorporate dissipation and thermalization to the
quantum map we consider the coupling of the system to a bath of
non interacting oscillators in thermal equilibrium. The degrees of
freedom of the bath are eliminated introducing the usual weak
coupling,  Markov and rotating wave
approximations \cite{QNoise}. This leads to a Lindblad equation
in action representation
\begin{equation}
\begin{array}{lll}
\!\!\!\!\!\!& \dot{\hat{\rho}} & =  \, -i [ \hat{H}_S, \hat{\rho}] \, + \, \\
\!\!\!& + & \! g \, \sqrt{n^+_{th} (\Omega_n,T)
n^+_{th}(\Omega_{n'},T)} \, \{ [\hat{L}_n, \hat{\rho}
\hat{L}_{n'}^{\dagger}] +
[\hat{L}_n \hat{\rho},\hat{L}_{n'}^{\dagger}]\} \\
\!\!\!& + & \! g \, \sqrt{n^-_{th} (\Omega_n,T)
n^-_{th}(\Omega_{n'},T)} \{ [\hat{L}_n^{\dagger}, \hat{\rho}
\hat{L}_{n'}] + [\hat{L}_n^{\dagger} \hat{\rho},\hat{L}_{n'}]\}
\end{array}
\label{qdiff}
\end{equation}
where the frequencies $ \Omega_n= n+1/2$ are the energy
differences between two neighboring levels of the rotator. The
population densities of the bath found in Eq. \ref{qdiff} are
given by
\begin{equation}
\begin{array}{l}
n^-_{th}(\omega,T)={1 \over exp(\hbar \omega /k_B T)-1} \\
\\
n^+_{th}(\omega,T)=n^-_{th}(\omega,T)+1
\end{array}
\label{population}
\end{equation}
The system operators $\hat{L}_n = \sqrt{|l|+1} (\ket{n} \bra{n+1}|
+ \ket{-n} \bra{-n-1}|)$ describe transitions towards the ground
state of the free rotator. Requiring quantum to classical
correspondence at short times we fix the coupling constant $ g = -
\ln (1-\Gamma)$. For $T=0$ we recover the master equation used in
\cite{Carlo} for the pure dissipative case. For finite
temperature, the last term in Eq. \ref{qdiff} describes the
thermal excitation of the rotator through absorption of heat bath
energy. Finally, Eq. \ref{qdiff} will be integrated numerically without 
further approximations.

\section{Results}

\begin{figure}[htp]
\begin{center}
\vspace{0.05\textwidth}
{\epsfxsize=8cm\epsffile{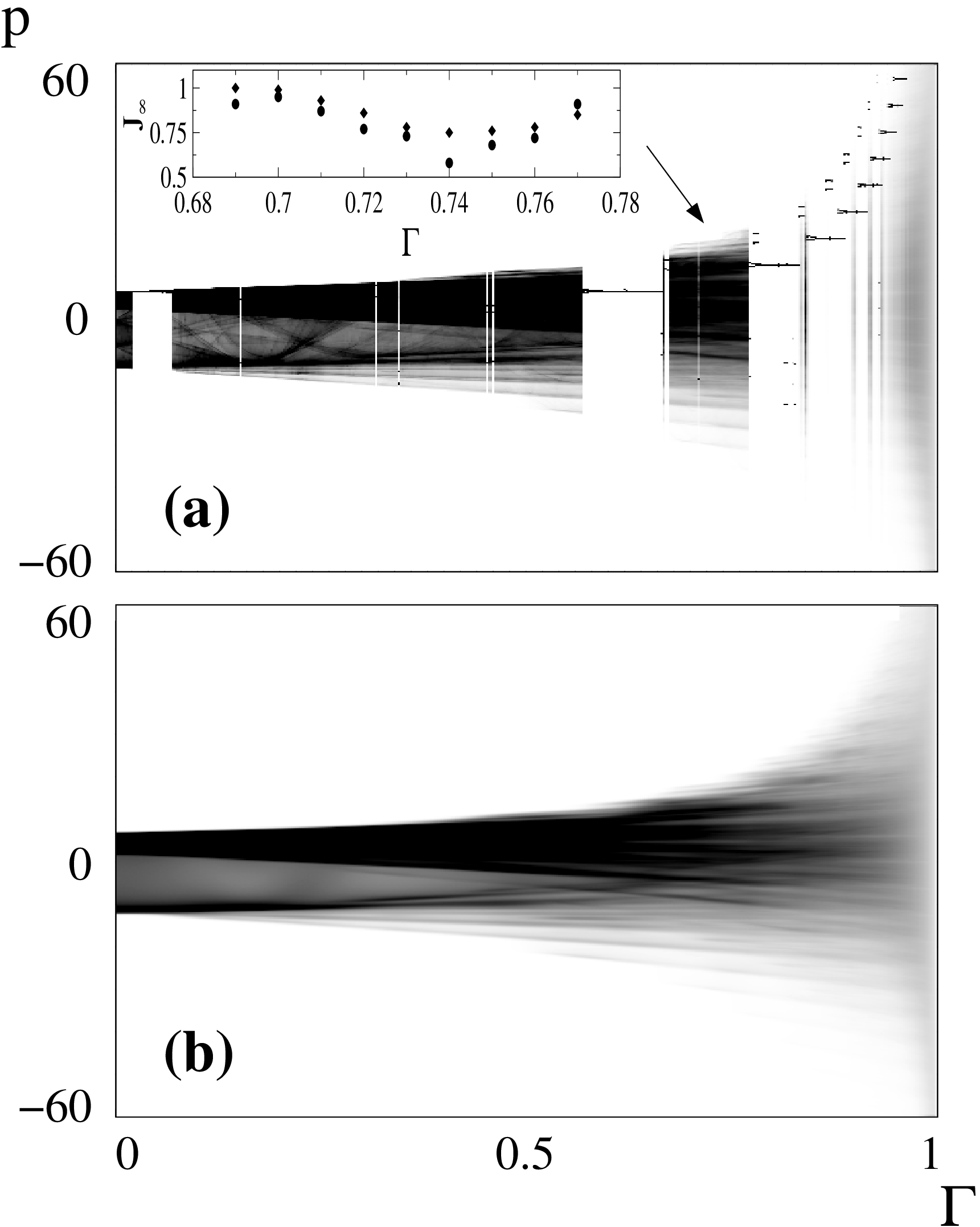}}
\caption{Bifurcation diagrams in terms of $p$, as a function of the parameter $\Gamma$.
We show the last $5\times10^3$ classical $p$
values corresponding to $5\times10^3$ random initial conditions taken in the interval
($x \in [0,\pi)$, $p \in [-\pi,\pi]$), and after $1.4\times10^5$ steps of the map. In panel (a) $T=0$
(inset shows the asymptotic current $J_{\infty}$ as a function of $\Gamma$ for the indicated irregular window),
while in (b) $T=0.05$.}
\label{fig:bifdiag}
\end{center}
\end{figure}

Since we are interested in chaotic transport, throughout the
following calculations we will use the set of parameters given by
$ K=0.7$, $\Phi= {\pi/2}$, and $a=0.7$. In the Hamiltonian limit,
this case shows no visible stability islands in phase space. We
have first studied some classical aspects of our system at zero
temperature, beginning with the bifurcation diagrams in terms of
$p$ as a function of the parameter $\Gamma \in [0,1]$ (see Fig.
\ref{fig:bifdiag}({\it a})). It should be mentioned that the
chaotic attractors set in very fast for small $\Gamma$. But this
is not the case for larger values where the transient times can be
very long. For this reason, we have calculated these diagrams with
the last $5 \times 10^3$ iterations of the map, after the first
$1.4 \times 10^5$ have been discarded. We have randomly taken $5
\times 10^3$ initial conditions inside the unit cell ($x \in
[0,\pi)$, $p \in [-\pi,\pi]$). Several regular and chaotic windows
alternate. The former are characterized by simple attractors
(stable fixed points of the dissipative map) and the latter are
dominated by chaotic attracting sets. The width in $p$ of these
sets grows as dissipation weakens ($\Gamma \rightarrow 1$).

The main quantity characterizing transport is the current
$J(t)=\langle p_t \rangle$, where $\langle \rangle$ stands for the
average taken on the initial conditions, and $p_t$ is the moment
after the $t$th iteration of the map.  Since bifurcations that
change the shape of strange attractors also play a role in
determining the values of the asymptotic current we restrict our
analysis to the chaotic window approximately located at $\Gamma
\in [0.68,0.78]$ indicated by an arrow. The inset of Fig.
\ref{fig:bifdiag}(a) displays the asymptotic current
$J_{\infty}=\lim_{t \rightarrow \infty} J(t)$ as a function of
$\Gamma$ (circles stand for $T=0$). At this $\Gamma$ range, 100
kicks are enough to reach the stationary behavior, independently
of the initial distributions . As pointed out in \cite{Carlo},
dissipation induces an asymmetry of the strange attractor which is
responsible for the directed transport. On the other hand this
dissipation mechanism contracts phase space and makes the higher
energies inaccessible for the system. The final value of $\langle p \rangle$
results from the interplay between both effects. In fact,
increasing dissipation does not increase the transport and the
largest values of $\langle p \rangle$ are obtained for the lowest values of
dissipation, i.e. $\Gamma \gtrsim 0.9$ (for example $J_{\infty}=5.78$ for
$\Gamma=0.97$). The minimum current in this window is reached for
an intermediate value of dissipation ($\Gamma=0.74$).

We then consider the case of finite temperature. The bifurcation
diagram corresponding to $ T=0.05 $ is shown in Fig.
\ref{fig:bifdiag}(b). It is clear that the effect of temperature
consists of smoothing the finer structure of the chaotic
attractors that is present for the smallest values of $\Gamma$.
Even for this extremely low value of $T$ the detailed features
have almost completely disappeared, with the exception of the
black lines corresponding to the highest values of the density
distributions. The other very interesting effect is that
temperature erases the regular windows allowing for a continuously
chaotic behavior. This could be of much relevance in obtaining
large ratchet currents without the need for an extremely fine
tuned, weak dissipation \cite{Casati} (this will addressed in 
future studies \cite{future}). As shown in the inset, low
temperatures (diamonds correspond to $ T=0.05 $) lead to a
noticeable enhancement of the asymptotic current $J_{\infty}$
(around $30\%$ for $\Gamma=0.74$).

\begin{figure}
\vspace{0.05\textwidth}
\centerline{\epsfxsize=8cm\epsffile{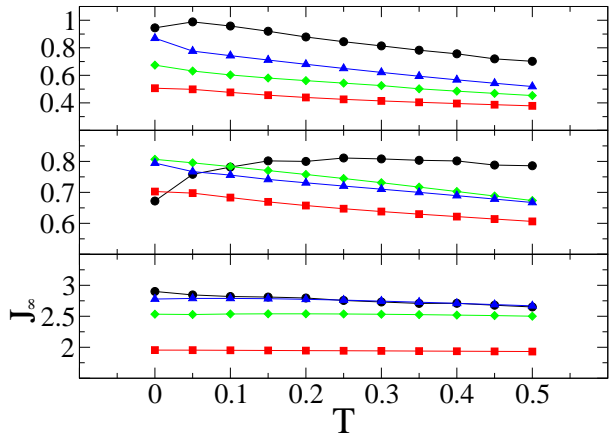}} \caption{(Color
online) Asymptotic current value $J_{\infty}$ as a function of
temperature $T$. Upper panel shows the case for $\Gamma=0.7$, the
middle one for $\Gamma=0.75$, and the lower one for $\Gamma=0.9$.
Black circles stand for the classical values. The quantum cases
correspond to $\hbar=0.055$ (blue triangles), $\hbar=0.165$ (green
diamonds), and $\hbar=0.494$ (red squares). As initial conditions
we have taken $10^6$ random points (classical) and a density
operator with equal population at all the possible $p$ eigenstates
inside the phase space region given by ($x \in [0,\pi)$, $p \in
[-\pi,\pi]$).} \label{fig:AllAsympCurrentsVsT}
\end{figure}

We now turn to compare the classical and quantum behaviors.
Firstly, we analyze the currents (which in the quantum case is
given by $J(t)=\tr{(\hat{\rho} \hat p)}$). In Fig.
\ref{fig:AllAsympCurrentsVsT} we display $J_{\infty}$ as a
function of $T$, for three different values of $\Gamma$ and
$\hbar_{\rm eff}$. At the classical level low temperatures lead to
an enhancement of the current for intermediate values of the
dissipation (see Fig. \ref{fig:AllAsympCurrentsVsT} upper and
middle panels corresponding to $\Gamma$= 0.7, 0.75) . In the case
of weak dissipation ( $\Gamma=0.9$ in the lower panel of Fig.
\ref{fig:AllAsympCurrentsVsT}), which displays larger values of
$J_{\infty}$, the effect of thermal noise is negligible. For
higher temperatures, the thermal effects reduce $\langle p \rangle$ in all
cases. This can be interpreted as follows: thermal noise reduces
the energy loss caused by dissipation (with no kicks, the system
would attain a Boltzmann distribution), so higher energies can be
reached, in comparison with a pure dissipative process. But since
this diffusion also tends to blur the asymmetry of the strange
attractor, the two effects compete and transport has a maximum for
low values of T, and then decreases.

At the quantum level we can clearly see that the previously
described thermal enhancement is not generally present, at least
for the $\hbar_{\rm eff}$ values we have considered. 
It is important to note that these values are consistent 
with experimental implementations using for example, cold sodium atoms 
in a laser field having a wave length $\lambda=589{\rm nm}$ 
(for more details see \cite{Carlo, KickedCA}). 
However, we observe a very slight growth of the current for the
$\hbar_{\rm eff}=0.055$ case with $\Gamma=0.9$ (see the blue
triangles in the lower panel of Fig.
\ref{fig:AllAsympCurrentsVsT})indicating that the temperature
dependence of the current is very sensitive to the particular
dynamics of the system. For a different example, an enhancement of the 
quantum transport has been observed \cite{Ghosh}, hence it is difficult 
to extract universal behaviors. For  $\Gamma=0.7$ the quantum current is lower than
the classical one for all temperatures, as already pointed out in
\cite{Carlo}, but only for $T=0$. The same happens in the case of weak damping
$\Gamma=0.9$ (nevertheless, for $\hbar_{\rm eff}=0.055$ and 
$T \gtrsim 0.1$ both currents coincide).  The case $\Gamma=0.75$ shows a different behavior.
At $T=0$ the quantum currents (for any of the $\hbar_{\rm eff}$ values we have considered) 
are larger than the classical ones, that is, there is an
enhancement due to the finite size of $\hbar_{\rm eff}$. Also, the maximal quantum current 
corresponds to $\hbar_{\rm eff}=0.165$, and this is valid for all the temperatures 
shown (for $T=0.05$ it is still greater than the classical one). For
larger quantum coarse-graining (see $\hbar_{\rm eff}=0.494$)
quantum currents decrease. In this sense, it seems that the effect
of quantum fluctuations on the quantum directed currents is
analogous to the effect of those of thermal origin on the
classical ones. Small fluctuations of thermal or quantum
mechanical origin assist directed transport while large
fluctuations (corresponding to high temperatures or to large
values of $\hbar_{\rm eff}$ respectively) blur the asymmetry of
the attractor and thus kill the net current.

It is interesting to note that in the case $\Gamma=0.75$ the
thermal diffusion associated with the temperature which gives
 the maximal value of the current ($  <\xi^2> =
0.12$ for $T= 0.25$) is of the order of the quantum
coarse-graining $\hbar_{\rm eff}$ corresponding to the strongest
quantum current. For $\Gamma=0.7$ the classical current attains its
maximum value at $T=0.05$ ($ <\xi^2> = 0.03$), which corresponds
to a value of $\hbar_{\rm eff}$ that we were not able to consider
in our numerical calculations.

\begin{figure}[htp]
\begin{center}
\vspace{0.05\textwidth} {\epsfxsize=8cm\epsffile{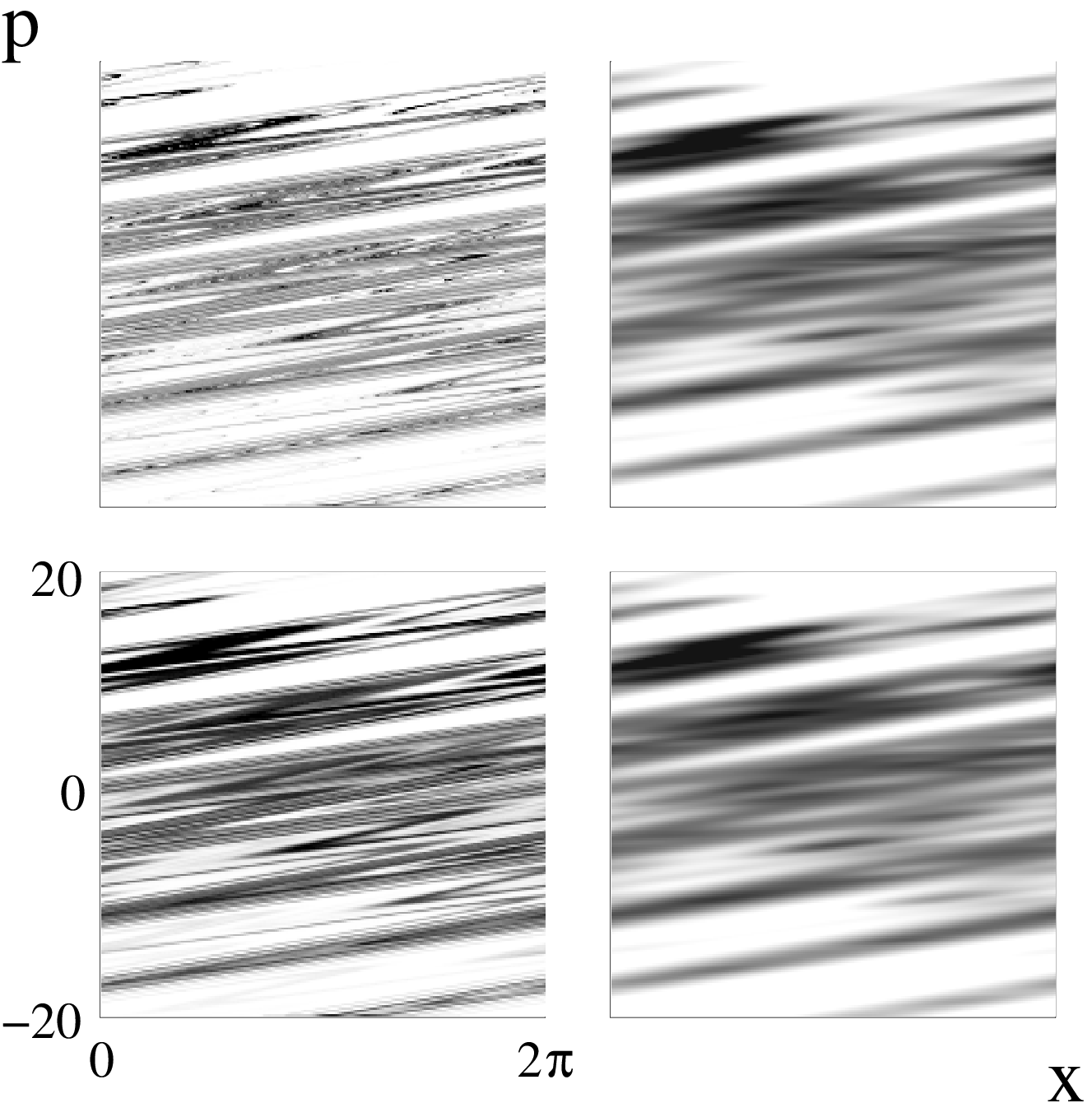}}
\caption{Phase space portraits for $\Gamma=0.75$ at $t=40$. Left
panels correspond to the classical while the right ones to the
quantum strange attractors. In the upper panels $T=0$, while in
the lower ones $T=0.1$. As initial conditions we have taken $10^6$
random points (classical) and a density operator with equal
population at all the possible $p$ eigenstates inside the phase
space region given by ($x \in [0,\pi)$, $p \in [-\pi,\pi]$).}
\label{fig:phasespace}
\end{center}
\end{figure}

The analogy between thermal noise and quantum coarse-graining can
also be appreciated when looking at the asymptotic Poincare
sections and Husimi distributions (displayed in Fig.
\ref{fig:phasespace} for $\Gamma=0.75 $, $\hbar_{\rm eff}= 0.055
$). As expected, at zero temperature the quantum Husimi function
reproduces well the main patterns of the classical attractor but
shows less fine structure (see the upper panels). If a small
temperature is introduced the fine details of the classical
distribution are smoothed out and both distributions look more
alike (see the lower panels corresponding to $T=0.1$). On the
other hand the quantum distributions at zero and finite
temperatures are practically indistinguishable, indicating that
the quantum coarse-graining is at least of the order of the
thermal one for these values of $T$.

\begin{figure}
\vspace{0.05\textwidth}
\centerline{\epsfxsize=8cm\epsffile{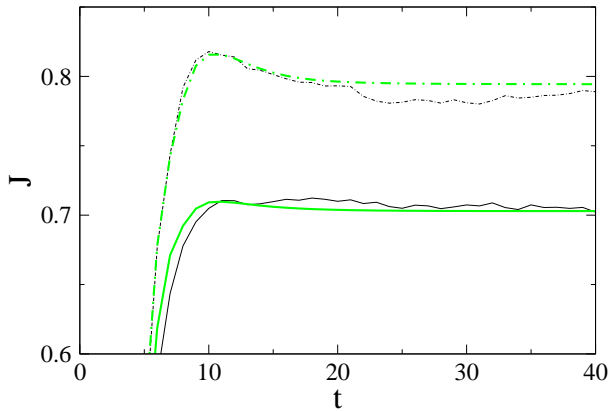}} \caption{(Color
online) Current $J$ as a function of time $t$, for coupling
strength $\Gamma=0.75$. Thin black lines correspond to the
classical values, while thick green (gray) ones to the quantum
cases (in these last cases we always take $T=0$). We show results
for $T=0.1$ and $\hbar_{\rm eff}=0.055$ (dot-dashed lines), and
for $T=0.85$ and $\hbar_{\rm eff}=0.494$ (solid lines). As initial
conditions we have taken $10^7$ random points (classical) and a
density operator with equal population at all the possible $p$
eigenstates inside the phase space region given by ($x \in
[0,\pi)$, $p \in [-\pi,\pi]$).} \label{fig:CurrentVstG075}
\end{figure}

We finally study $J(t)$ as a function of $t$ (i.e., the number of
iterations of the map). Results for $\Gamma=0.75$ are shown in
Fig. \ref{fig:CurrentVstG075}, where different temperatures and
$\hbar_{\rm eff}$ values have been considered. Besides the
mentioned fact that the asymptotic value is reached very rapidly,
we notice that the transient behavior shows a very close
quantum-to-classical correspondence. The classical current peak
observed at $t \sim 10$ for low temperatures ($T=0.1$) is also
present in the quantum current when $\hbar_{\rm eff}=0.055$ and
$T=0$. This peak disappears from the classical current at larger
temperatures ($T=0.85$), and so does the quantum one at
$\hbar_{\rm eff}=0.494$ and $T=0$. So the analogy seems to hold at
all times.

\section{Conclusions}

In this work we have analyzed the behavior of a chaotic dissipative system that shows
directed transport under the influence of a thermal bath, both in its classical
and quantum versions. We have varied the strength of the coupling with the environment
and also the temperature. We have found that the transport enhancement effect due to a finite 
temperature is highly dependent on the system specific properties. In fact, it depends on the 
coupling strength of the system with the environment and also on the $\hbar$ size.
Moreover, we could find an analogy between the effects caused by thermal and
quantum fluctuations. These results open the possibility for many further studies that
include finding ways of obtaining large ratchet currents in experimentally realistic
situations in kicked BECs and cold atoms experiments. These are one of the best candidates
to test our results since even BECs show an unavoidable fraction of noncondensed atoms 
when kicked. If kicks become strong, thermal excitations will be of much relevance rather 
than a negligible effect. With the sort of calculations presented in this paper the 
effects of this fraction on the transport properties of the system could be estimated.

\begin{acknowledgments}
Partial support by ANPCyT and CONICET is gratefully acknowledged.
\end{acknowledgments}

\end{document}